# *Ab initio* study of the formation of transparent carbon under pressure

**Xiang-Feng Zhou (周向锋),[1] Xiao Dong (董校),[1] Guang-Rui Qian (钱广锐),[3]**

**Lixin Zhang (张立新),[1] Yonjun Tian (田永君),[2] and Hui-Tian Wang (王慧田)[1,3]**

[1]School of Physics and Key Laboratory of Weak-Light Nonlinear Photonics, Ministry of Education, Nankai University, Tianjin 300071, China

[2]State Key Laboratory of Metastable Materials Science and Technology, Yanshan University, Qinhuangdao 066004, China

[3]Nanjing National Laboratory of Microstructures, Nanjing University, Nanjing 210093, China

**Abstract**

A body-centered tetragonal carbon (bct-Carbon) allotrope has been predicted to be a transparent carbon polymorph obtained under pressure. The structural transition pathways from graphite to diamond，M-Carbon, and bct-Carbon are simulated and the lowest activation barrier is found for the graphite-bct transition. Furthermore, bct-Carbon has higher shear strength than diamond due to its perpendicular graphene-like structure. Our results provide a possible explanation for the formation of a transparent carbon allotrope via the cold compression of graphite. We also verify that this allotrope is hard enough to crack diamond.

Carbon exists in a large number of forms thanks to its ability to form sp- , $sp^2$-, and $sp^3$-hybridized bonds, creating graphite, hexagonal diamond (lonsdaleite), diamond, nanotubes, fullerenes, and amorphous carbon. [1-7] The cubic diamond phase of carbon remains the hardest known solid at room temperature. Because of the extensive applications of diamond, intense theoretical and experimental efforts have been devoted to searching for materials that have comparable or even higher hardness and thermal stability. [8-9] For instance, some polycrystalline samples transformed from graphite under high pressure and temperature has been reported, the products of which feature equal or higher hardness than single crystal diamonds. [9]

Recent cold compression experiments have indicated that possible carbon polymorphs exhibit exceptionally high indentation strength, sufficient to indent diamond anvils. [4-5] Some samples were proven to be quenchable at room temperature, [4] while others were not.[5] Graphite, in particular, is an ultrasoft material under ambient conditions due to its weak van der Waals interactions among the interlayers. However, under cold compression, it shows substantial shear strength enhancement. [5]

This unexpected enhancement in indentation strength raises many fundamentalproblems regarding the exact crystal structures during phase transition and the nature of their characteristics. Hexagonal diamond, an intermediate or modified hexagonal phase between graphite and diamond, or an amorphous phase, were originally considered.[1,2,10] Most recently, Li *et al.* using the *ab initio* evolutionary algorithm, found that a mixture of graphite and M-Carbon could better explain the x-ray diffraction (XRD) patterns and near K-edge spectra obtained. [11]

In this work, an *ab initio* pseudopotential density functional method code within the local-density approximation (LDA) as implemented in the **CASTEP** is employed to carry out first-principles calculations.[12] Norm-conserving pseudopotentials are used in conjunction with plane-wave basis sets of cutoff energy of 660 eV and a Monkhorst-Pack Brillouin zone sampling grid spacing of 0.04 $Å^{-1}$. The electron-electron exchange interaction is described by the exchange-correlation function of Ceperley and Alder, as parameterized by Perdew and Zunger. [13] During the geometry optimization, neither symmetry nor restrictions are constrained for either the unit cell shape or the atomic positions with respect to the Broyden-Fletcher-Goldfarb-Shanno (BFGS) minimization scheme. The structural relaxation is stopped when the total energy, the maximum ionic displacement, the maximum stress, and the maximum ionic Hellmann-Feynman force are less than $5 \times 10^{-6}$ eV/atom, $5 \times 10^{-4}$ Å, 0.02 GPa, and 0.01 eV/ Å, respectively. To obtain more accurate band gaps, the hybrid functional (according to Becke's exchange functional), combined with the Lee-Yang-Parr correlation (B3LYP) are used. [14,15]

The tensile and shear stress are computed as follows: The desired target-stress component is set to a certain value while other components are kept zero. The lattice vectors and atomic positions are then relaxed simultaneously to obtain the final structures. We increase the desired target-stress component step by step and repeat the above procedure until the structure collapses, in which case the maximum stress is considered to be the ideal strength. [16-19]

Routine first-principles calculations are performed to clarify the high-pressure phases of carbon. The structure search involves relaxing a set of randomly chosen structures or modifying ones constructed from various orientations of prototype crystals, until the energy arrives at its minimum at ground state or a given pressure. [20-23] This results in a carbon phase with body-centered tetragonal I4/mmm symmetry, as shown in Figs. 1(a) and 1(b), designated as bct-Carbon. This structure appears to be similar to that found in a previous study using molecular dynamic simulations of carbon nanotubes at 20 GPa.[24] Its lattice parameters were a = b = 4.322 Å and c = 2.478 Å, respectively. There is a nonidentical C atom occupying the 8h (0.18, 0.18, 0) site. By comparing simulated XRD patterns obtained from this product to those of M-Carbon and experimental data, bct-Carbon is confirmed to be a viable candidate for transparent and superhard graphite under cold compression. [25-27] Table I lists some parameters for bct-Carbon, M-Carbon, and diamond at ambient pressure. Two different kinds of C-C bonds exist in bct-Carbon, with bond lengths of 1.559 and 1.503 Å at equilibrium. The average bond length is 1.531 Å, which is comparable with 1.533 Å for M-Carbon (there are 8 kinds of C-C bonds, Table I lists the minimum and maximum values only) and 1.545 Å for diamond. The results imply that bct-Carbon and M-Carbon should have bond strengths similar to that of diamond. [28]

To reveal the formation mechanism of the transparent carbon allotrope, a variable-cell nudged elastic band (VC-NEB) method is developed to obtain the energy barrier and transition paths for bct-Carbon, M-Carbon, and diamond (the details are described in the supplementary material [29]). The present results show the bidirectional energy barriers to be 0.13 eV/atom from graphite to bct-Carbon and 0.15 eV/atom from bct-Carbon to graphite, as shown in Fig. 2, which are lower than 0.15 eV/atom from graphite to diamond and 0.43 eV/atom for the reverse process. Such a minor difference of 0.02 eV/atom between the forward energy barrier and the backward one for bct-Carbon, compared with 0.28 eV/atom for diamond, means that bct-Carbon will be “easy-come-easy-go” under pressure, that is to say, bct-Carbon will be easier to be formed from graphite, compared with diamond from graphite, but may not be quenchable at room temperature. This is also consistent with experimental observations. [1, 5] The transition path of bct-Carbon is carefully verified. The results indicate that the hexagonal graphite first transforms to rhombohedral graphite under a pressure of 20 GPa. The pressure-induced bonding instability in the rhombohedral graphite then leads to significant bond-length fluctuations for the intra-layer of graphite (see the inset of Fig. 2 or Fig. 1 of Ref. 29), which dominates the energy barrier of the phase transition. These fluctuating graphene-like structures are further compressed, ultimately forming a “4+8” structure, shown in Fig. 1(b). Therefore, the special crystallography of bct-Carbon also provides direct theoretical evidence of Mao's prediction [see Fig. 2 in Ref. 5], where, under compression, bridging carbon atoms pair with other atoms in the adjacent layers to form the $\sigma$ bonds. The residual $\pi^*$ components in the near K-edge spectrum arise from the incomplete conversion of graphite. [11] For M-Carbon, the forward and backward energy barriers are 0.28 and 0.35 eV/atom, respectively. The results show that the bidirectional energy barriers of M-Carbon are higher than those of bct-Carbon, while the backward energy barrier is lower than that of diamond. By carefully checking the detailed transition pathway, M-Carbon exhibits a similar compression behavior to bct-Carbon. [29] The distinct difference in the transition path is attributed to the fluctuating behavior (up and down) and degree (large or small) of the graphene layers. The relatively large fluctuation of graphene layers in M-Carbon (see Fig. 2 of Ref. 29) with respect to bct-Carbon is one of the dominant reasons for the large enhancement in energy barrier; that is to say, M-Carbon seems to be undulated larger than that of bct-Carbon during phase transition. These fluctuating graphene-like structures are further rolled, ultimately forming a "5+7" structure instead of a "4+8" structure for bct-Carbon. [11, 29, 30] We should also note that our present results are based on the theoretical prediction within the NEB method. Transition barriers are sensitive to the computational method, and the general methodological question on which method is best is not yet settled. Using other methods, such as molecular dynamics or transition path sampling, one might obtain different results for the energy barrier between bct-Carbon and M-Carbon. While it is too early to make a definitive conclusion that bct-Carbon will be superior than M-carbon as the best candidate for the cold-compressed graphite, we can confidently say that in both cases the transition must involve puckering of the graphene sheets with the formation of covalent bonds between the layers. Further studies of the transition barrier with other methods are highly desirable for complete understanding of the high-pressure behavior of carbon. Our results are the important step in this direction and show that bct-Carbon is kinetically easier to form than other carbon allotropes within the NEB method. Since bct-Carbon is still not unambiguously identified from experiment, one should be very careful to compare our results directly with experimental values.

We now focus on the superior indentation strength of graphite under cold compression. By careful choosing the applied deformation, nine elastic constants can be determined. Combined with Voigt-Reuss-Hill approximation, [31] the calculated bulk and shear modululi of 447 and 540 GPa for diamond are in good agreement with the respective experimental values of 442 and 544 GPa. [32] Using the same method, the bulk and shear modululi of bct-Carbon are calculated to be 414 and 427 GPa, respectively, both of which are lower than those of diamond. Therefore, bct-Carbon is not harder than diamond, but is similar to M-Carbon (415 and 468 GPa).

The bulk and shear modululi are not necessary to give accurate accounts of the strength of a material. This is because these elastic constants are evaluated at the equilibrium state, whereas material deformations associated with cold compression measurements usually involve large strains where bonding characteristics may substantially change. [18] Thus, the ideal strength calculation may be a good alternative method for estimating the indentation strength. Diamond is first tested, as shown in Fig. 3. The calculated tensile and shear strengths along the weakest direction are 91.1 and 92.5 GPa at strains of 0.13 and 0.27, respectively, which are consistent with results reported in literature. [16-19] These results are then compared with those obtained for bct-Carbon. In the cases of tensile load along the [100], [010], [001], [110], and [111] directions, the corresponding tensile stresses are found to be 84.8, 84.8, 139.7, 131.1, and 107.5 GPa, respectively. Clearly, in both [100] and [010] directions, the tensile stresses are equal because the two directions are identical, as shown in Fig. 1(b). Consequently, the ideal tensile strength of bct-Carbon is 84.8 GPa, which is lower than the value 91.1 GPa found for diamond.

Since the shear strength is closely related to the indentation strength, the shearing caseis explored and verified. The shear strengths in the (001) [100], (001) [010], (010) [001], (010) $[00\bar{1}]$, and (010) [100] systems are found to be 108.6, 108.6, 108.6, 108.6, and 119.7 GPa, respectively. Evidently, the shear strengths in the first four directions are weakest and identical, implying that the weakest slip systems are in these directions. Therefore, the ideal shear strength of bct-Carbon should be 108.6 GPa, which is larger than the shear strength of diamond (92.5 GPa) by at least 17%. The inset in Fig. 3 shows a snapshot of bct-Carbon at a shear stress of 108.6 GPa. The weakest C-C bond (1.669 Å) will not break at a relatively large strain of 0.29, which denotes substantial endurance beyond the linear elastic regime. Thus, the exceptional shear strength may be understood by the conception of *crystallographic strength*, [33] because bct-Carbon has a perpendicular graphene-like structure, as shown in Fig. 1(a). Graphene is the hardest material with the strongest intrinsic bond strength. [34] Therefore, a perpendicular graphene-like configuration is expected to be able to withstand larger critical stresses from different directions and retard the occurrence of weak slip systems, as well as any instability towards graphite under cold compression.

In summary, we have demonstrated the existence of a pressure-induced bct-Carbon phase with body-centered tetragonal I4/mmm symmetry. This possible phase may be obtained from graphite under cold compression. The calculated transition pressure and simulated XRD pattern are in good agreement with the results from experimental observations. [26, 27] The transparent carbon allotrope, bct-Carbon, has a broad band gap similar to that of diamond. [27] Although the most likely transition path for this allotrope has a lower activation barrier compared with those of M-carbon and diamond within NEB calculations, we deem it most likely that coexistence of bct-Carbon and M-Carbon may occur in the cold-compressed graphite under pressure. [26] It also should be emphasized that the perpendicular graphene-like structure of bct-Carbon is responsible for its exceptional shear strength, which is large enough to crack diamond.

This work was in part supported by the National Natural Science Foundation of China (Grant No. 50821001), by the 973 Program of China (Grant Nos. 2006CB921805, 2005CB724400, and 2011CB808205), and the Postdoctoral Fund of China (Grant No. 20090460685).

## Figure Captions

Fig. 1 (a) and (b) Views along [100]/[001], and [010] directions of $2\times2\times2$ supercell of bct-Carbon，the dotted-dashed in (b) indicate the perpendicular graphene-like structure of bct-Carbon.

Fig. 2 Energy barrier curves of bct-Carbon (blue squares), M-Carbon (green triangles), and diamond (red circles) at pressure of 20 GPa. The inset shows the fluctuant graphene layers from the intermediate image of the transition path which is responsible for the activation energy barrier.

Fig. 3 The calculated stress-strain relations of bct-Carbon compared with that of diamond. The filled (hollow) blue square and red circle are used to depict the tensile and shear stress of bct-Carbon and diamond, respectively. The arrows indicate the ideal strength (including tensile and shear strength). The arrows indicate the ideal strength (including tensile and shear strength). The inset shows the electron density of bct-Carbon in the (010) plane (units $Å^{-3}$) at largest strain, in which the white dotted lines imply that new graphene layer will be reformed in this range as the shear stress or strain increases.

Table I: Structural parameters, coordination, bond length, volume, and band gap for bct-Carbon and M-Carbon (only the maximum and minimum value are listed for M-Carbon) compared with those of diamond at zero pressure.

| Polymorph | Coordination | Bond length (Å) | Volume (Å$^3$/atom) | Band gap (eV) |
|---|---|---|---|---|
| bct-Carbon | 4 | 1.503 (1.506[a])<br>1.559 (1.562[a]) | 5.787 (5.82[a]) | 5.14 (3.78[a]) |
| M-Carbon | 4 | 1.484 (1.489[b])<br>1.604 (1.607[b]) | 5.739 (5.78[b]) | 5.16 (3.6[b]) |
| Diamond | 4 | 1.545<br>Expt. 1.54[a] | 5.478<br>Expt. 5.68[a] | 5.85<br>Expt. 5.5[c] |

[a] Reference [26]. [b] Reference [11]. [b] Reference [35].

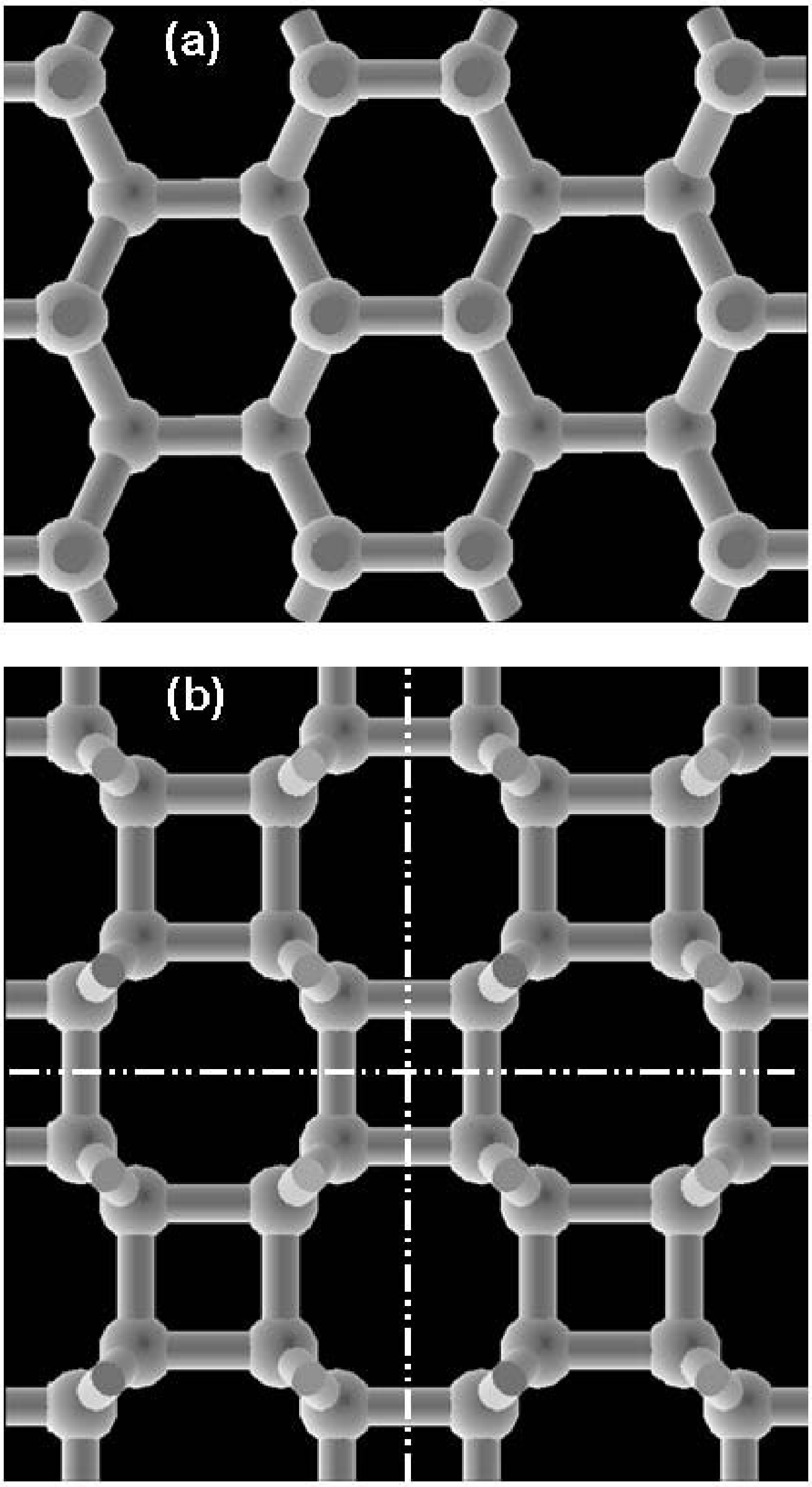

Fig. 1

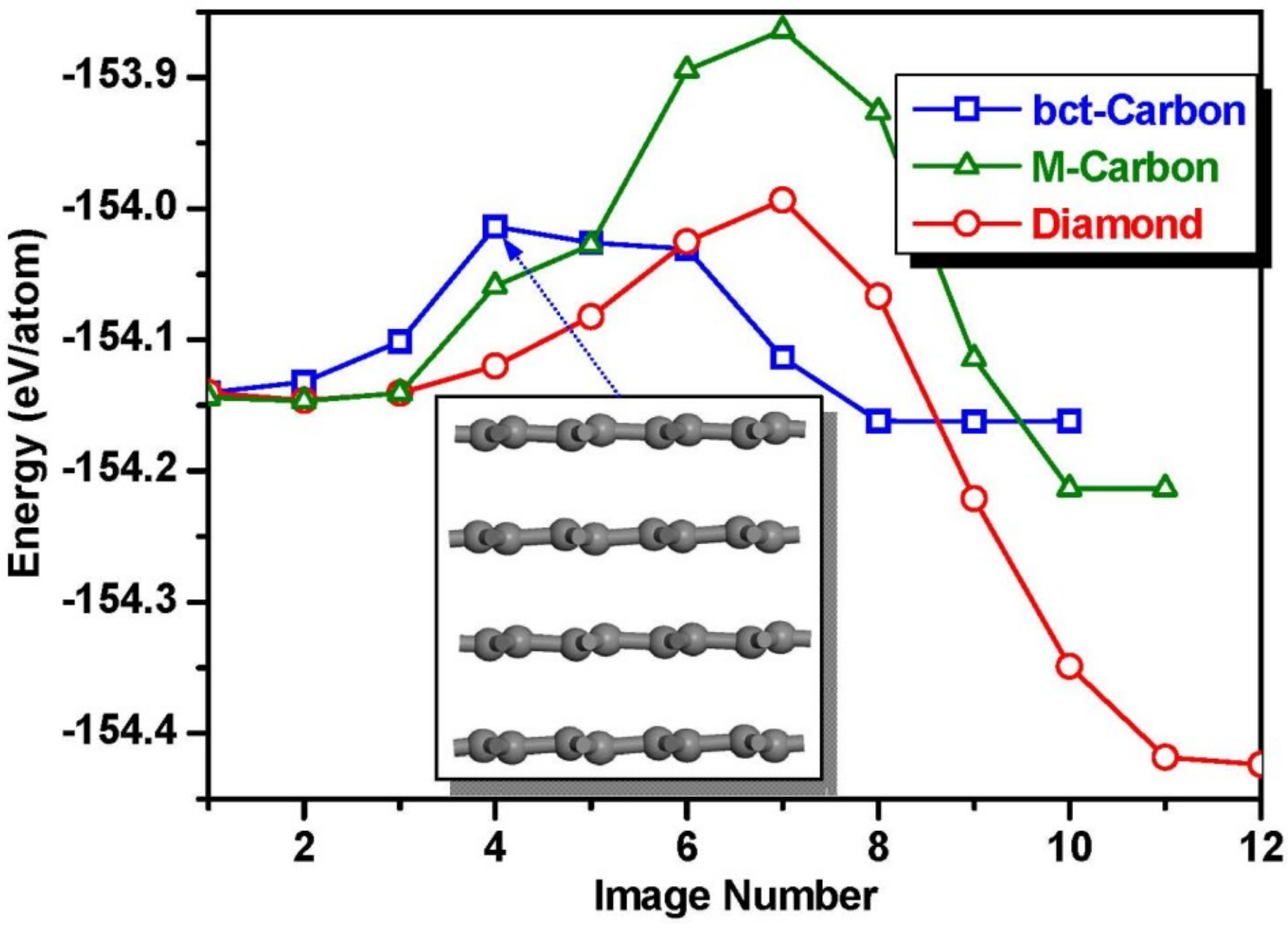
-153.9
-154.0
-154.1
-154.2
-154.3
-154.4
Energy (eV/atom)
2
4
6
8
10
12
Image Number
bct-Carbon
M-Carbon
Diamond

Fig. 2

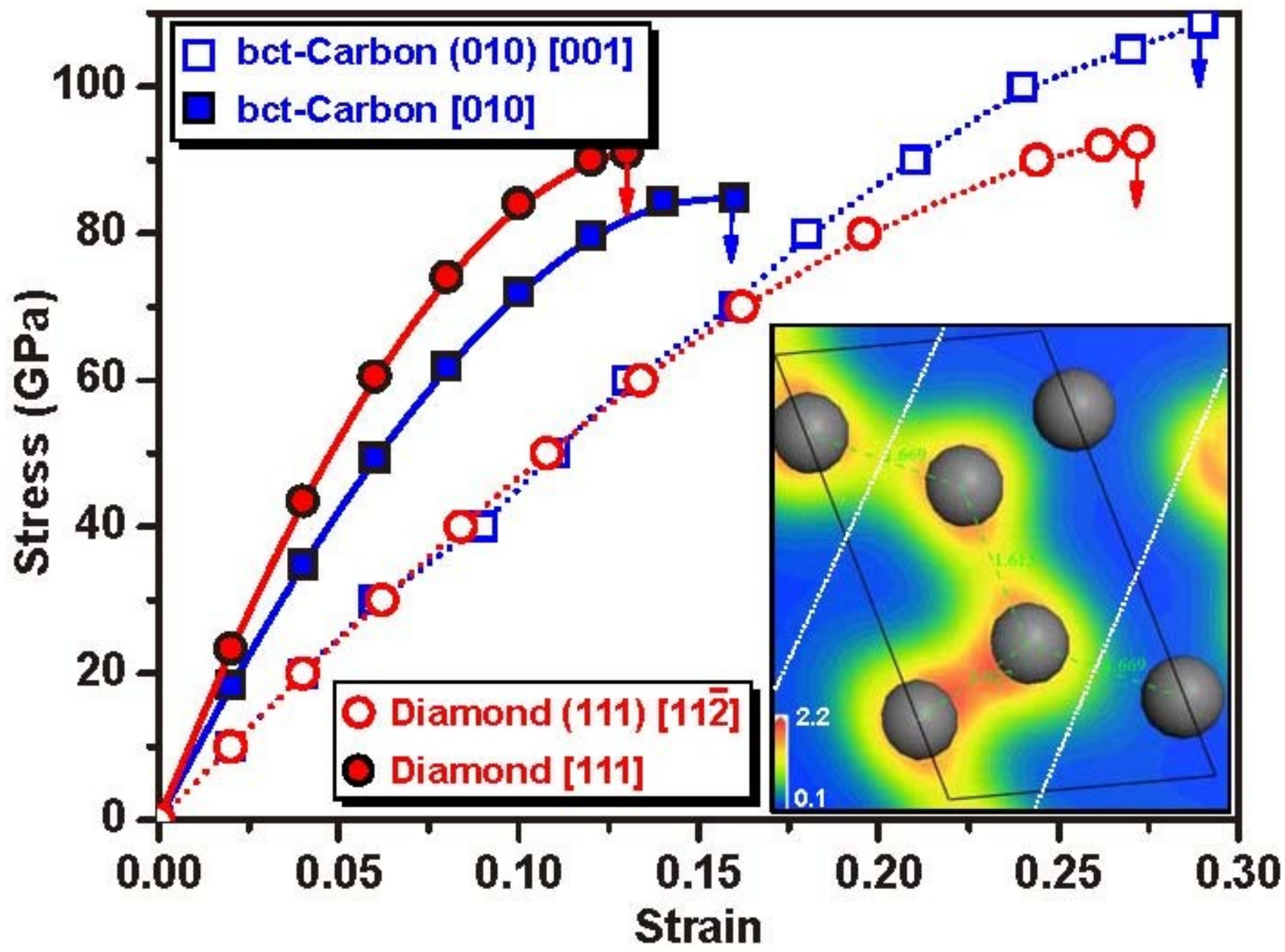
bct-Carbon (010) [001]
bct-Carbon [010]
Diamond (111) [11$\bar{2}$]
Diamond [111]
Stress (GPa)
Strain
0
20
40
60
80
100
0.00
0.05
0.10
0.15
0.20
0.25
0.30
2.2
0.1

Fig. 3

**Supplementary Online Material**

## Additional DFT calculations

As a widely used method, nudged elastic band (NEB) [1-3] method is an efficient and successful approach for finding the reaction pathways and the saddle points along the "minimum energy path" (MEP) between the two endpoints. It has been successfully applied in the realm of chemical reactions of molecules, metal surface, and defect migration, for estimating the activation energy barrier between the given initial and final states of a transition. We developed a variable cell nudged elastic band (VC-NEB, compared with the traditional fixed cell NEB) method [4], for extending to a constant pressure condition combining with the variable cell approach [5]. The VC-NEB method, which includes the unit cell deformation, provides a broad way to find MEP and investigates the activation pathways between the two given phases for a phase transition process within a larger configuration space. For all the VC-NEB calculations in this work, we take many intermediate images besides the two endpoint phases and choose the force and energy convergences to be the levels of 0.01 eV/Å and 0.001 eV, respectively. The variation of the spring constant is restricted within the range of 0.2-1.5 eV/Å$^2$, which is slightly narrower/tighter than that suggested in Ref. 1. All the calculations were implemented under the density functional theory framework using the PWSCF code [6]. To confirm the reliability of our method, we investigated many samples extensively. However, here we showed the energy barrier of zinc oxide (ZnO) under pressure, and compare them with those of previous published results. As is well known, there are two most likely paths (hexagonal or tetragonal path) for the phase transition from wurtzite to rocksalt ($B_4 \rightarrow B_1$) [7-10]. We used the generalized gradient approximation (GGA) for the exchange correlation functional, ultrasoft pseudopotential, and based on convergence tests adopted a kinetic energy cutoff of 75 Ry and a 8 × 8 × 6 MP mesh for Brillouin zone (BZ) integration. The energy barrier for the tetragonal path of ZnO is 0.134 eV/formula, compared with 0.132 eV/formula for the hexagonal path which is in good agreement with previous results (~0.15 eV/formula) [11]. Such minor difference for the energy barrier between the two different paths (0.002 eV/formula) implies the coexistence of the two independent paths in the phase transition at the same time, which is also consistent with the results found by molecular dynamics calculation [12].

For the high pressure phase of carbon, the energy landscapes will be much complicated than that of ZnO. For example, the transformation of graphite to new (unknown up to date) superhard phase at cold compression is not very clear at present. M-Carbon or bct-Carbon is predicted to be the most likely candidates at present. However, cubic or hexagonal diamond will be much more stable than those of M-Carbon or bct-Carbon under the same pressure. It increases the difficulty greatly to find the energy barrier exactly because the variable cell method often tries its best to find the transition path which includes the potential most stable configurations. To get the energy barrier of bct-Carbon and M-Carbon under pressure, we used the local density approximation (LDA) for the exchange correlation functional, norm-conserving pseudopotential, an 80 Ry pane-wave cutoff energy, and a $6 \times 6 \times 6$ MP mesh for the k point sampling of BZ. At first, the energy barrier from graphite to diamond under pressure of 20 GPa is 0.15 eV/atom, which agrees satisfactorily with the published results [13]. It should be noted that the energy barrier will be changed with external pressure and it is indeed a very time-consuming work. We simulated the transition paths of bct-Carbon and M-Carbon from different configurations, and at least got almost the same energy barriers from two most likely independent transition path ways. Figures 1 and 2 list the detailed transition path for bct-Carbon and M-Carbon, respectively.

## Figures

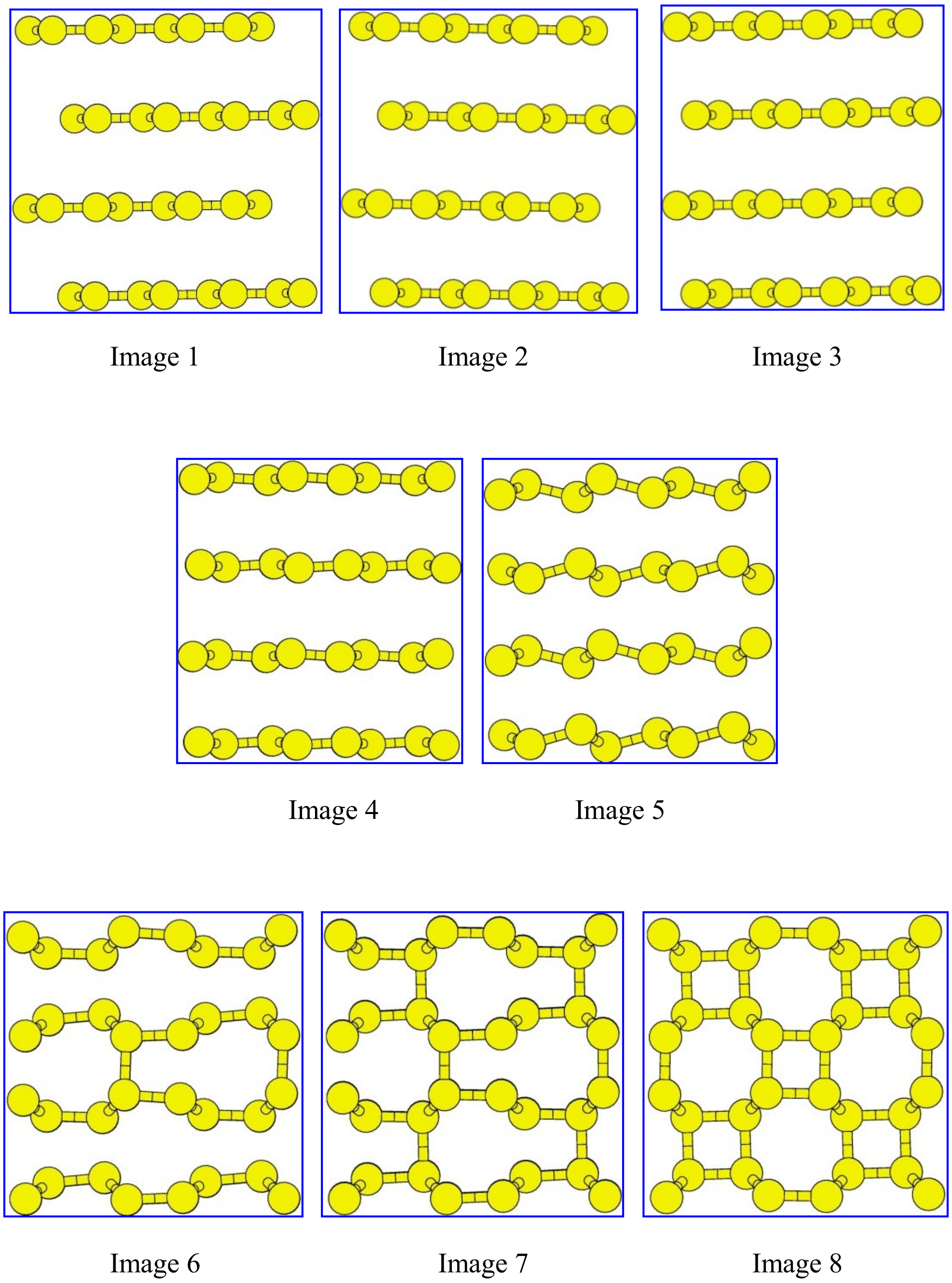

Fig. 1 Detailed transition path of bct-Carbon under pressure of 20 GPa. The sideview of structures at different intermediate images shows there is a continuous pathway between graphite and bct-Carbon.

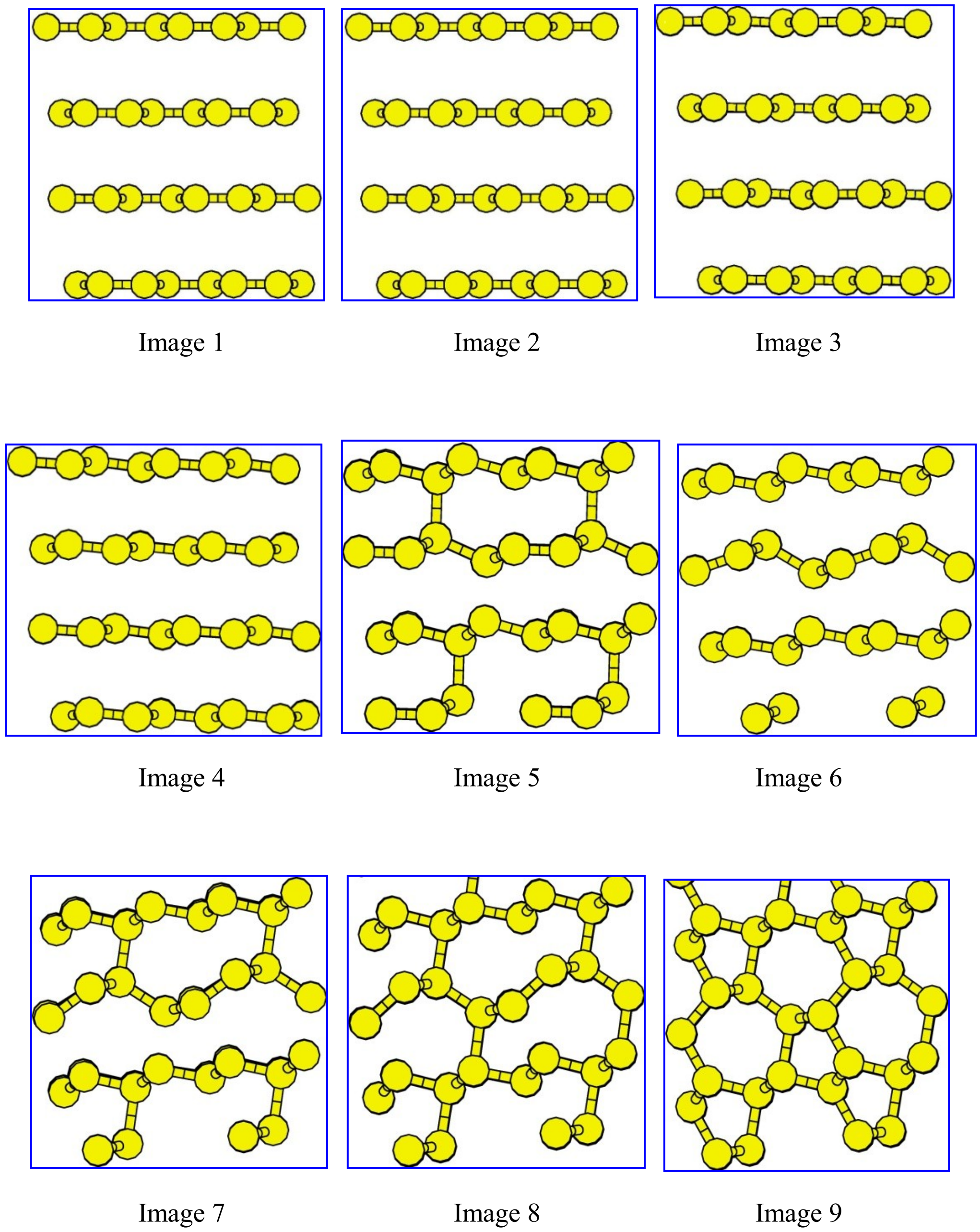

Fig. 2 Detailed transition path of M-Carbon under pressure of 20 GPa. The sideview of structures at different intermediate images shows there is a continuous pathway between graphite and M-Carbon.